\documentstyle[epsf]{mn}

\title[Clustering of Galaxies at High Redshifts]
      {Clustering of Galaxies at High Redshifts}
\author[J.S.Bagla]
       {J.S.Bagla\\
        Institute of Astronomy, University of Cambridge, Madingley Road,
        Cambridge CB3 0HA, U.K.\\
        E-mail: jasjeet@ast.cam.ac.uk}

\pagerange{\pageref{firstpage}--\pageref{lastpage}}
\pubyear{1997}

\begin{document}
\label{firstpage}

\maketitle

\begin{abstract}
Recent observations show a large concentration of galaxies at high
redshift.  At first sight strong clustering of galaxies at high
redshifts seems to be in contradiction with the models of structure
formation.  In this paper we show that such structures are a
manifestation of the strong clustering of rare peaks in the density
field.  We compute the frequency of occurrence of such large
concentrations of galaxies in some models of structure formation.
\end{abstract}


\section{Introduction}

Observations of galaxies at high redshifts, around $z=3$, in a
contiguous region of sky \cite{wallobs} show a marked concentration in
a region of width $\Delta z =0.04$, or $\Delta v =
3000$km s$^{-1}$.  The angular size of the region probed is $9'
\times 18'$ and within this region the galaxies contributing to the
peak are distributed randomly in about half the area.  A quasar is
also present in this wall like structure.  The clustering in redshift
space is estimated to have a confidence level of $99.8\%$.  Steidel at
al. (1997) point out that clustering at such large scales, in most
models of galaxy formation, requires high bias to explain the observed
structure. 

First galaxies to form in the universe correspond to the deepest
potential wells or the highest peaks in the initial density
distribution.  It is well known that these rare peaks cluster more
strongly than, say, the typical peaks \cite{bbks}.  Therefore, at
early times, when only rare peaks have collapsed into structures like
galaxies, we expect these objects to show significant clustering.
The epoch when typical halos of a given mass scale $M$ collapse is
characterised by $\sigma(M,z) \simeq \delta_c$ \cite{ps75}.  It is
customary to use $\delta_c = 1.69$ as a spherical perturbation
virialises when the linearly extrapolated density contrast equals
$\delta_c$ \cite{gunngott}.  If we define a quantity
$\nu(M,z)=\delta_c/\sigma(M,z)$, then we can say that typical halos of
mass $M$ collapse when $\nu(M) \simeq 1$.  For $\nu \gg 1$, only the
rare peaks of the given mass scale have collapsed and these tend to
cluster very strongly.  In this paper, we study clustering of halos at
$z \approx 3$ and compare it with the constraint set by the
observation of Steidel et al. (1997) for some models of structure
formation.  

We assume, that the galaxy distribution and the halo distribution
are identical.  This may not be true, especially at small scales where
gastrophysical processes play an important role.  However, including
these processes requires more detailed modelling which, in turn,
requires more assumptions.  Therefore, we will restrict the present
study to the study of halo distribution at high redshifts.

We address the following questions here:
\begin{itemize}
\item  Is the clustering of dark matters halos in models of structure
  formation sufficiently strong to   explain the observed
  concentration of galaxies? 
\item  What is the frequency with which we may expect to see a
  $3\sigma$ excess for the number of halos if we simulate the
  observations of galaxies using numerical simulations?
\end{itemize}

\section{Clustering of Halos}

We can study the clustering of halos using numerical simulations.  For
this particular study we choose the SCDM model with the shape
parameter $\Gamma = 0.5$ \cite{ebw}.  We normalise the power spectrum
so that $\sigma_8=\sigma(8 $h$^{-1}$Mpc$, z=0)=0.6$.  Most models have
more power at the relevant scales as compared to this, and if this
model provide enough clustering at $z=3$, most other models can do
that too.

The observed number density of Lyman break galaxies is $2.9\times
10^{-3}$h$^3$Mpc$^{-3}$ in the Einstein-de Sitter Universe.  This
implies a halo mass of about $5\times 10^{12}M_\odot$ for the SCDM
model \cite{ps75}.  Using this mass for halos is same as assuming that
all halos host Lyman break galaxies.  In order to relax this
assumption, we choose a lower halo mass, $M_{halo} \ge 7\times
10^{11}M_\odot$.  Thus less than one in ten of these halos host Lyman
break galaxies.  In absence of any alternative, we assume that
galaxies are distributed randomly amongst these halos and hence have
the same clustering properties.

We quantify the clustering of halos with the averaged two point
correlation function.  This is defined as
\begin{equation}
\bar\xi(r) = \frac{3}{r^3} \int\limits_0^r x^2 \xi(x) dx = \frac{3
  J_3(r)}{r^3} 
\end{equation}
where $\xi(x)$ is the two point correlation function.  

Halos are identified using the Friends-of Friends (FOF) algorithm with
a linking length of $0.2$ (over-density of $60$).  Fig.1 shows the
averaged correlation function as a function of scale 
for these halos at redshift $z=3$.  If the number density
of these halos is $\bar n$, then the number of halos within distance
$r$ from a typical halos is $4 \pi r^3 {\bar n}(1 + \bar\xi(r))/3$.
Thus we expect to see twice the average number in a sphere of radius
$8$h$^{-1}$Mpc around a typical halo.  Some halos will have more
neighbours than a typical halo.  Therefore, an excess of a factor four
in one bin is not very surprising.

\begin{figure}
\epsfxsize=3.3truein\epsfbox[39 26 502 494]{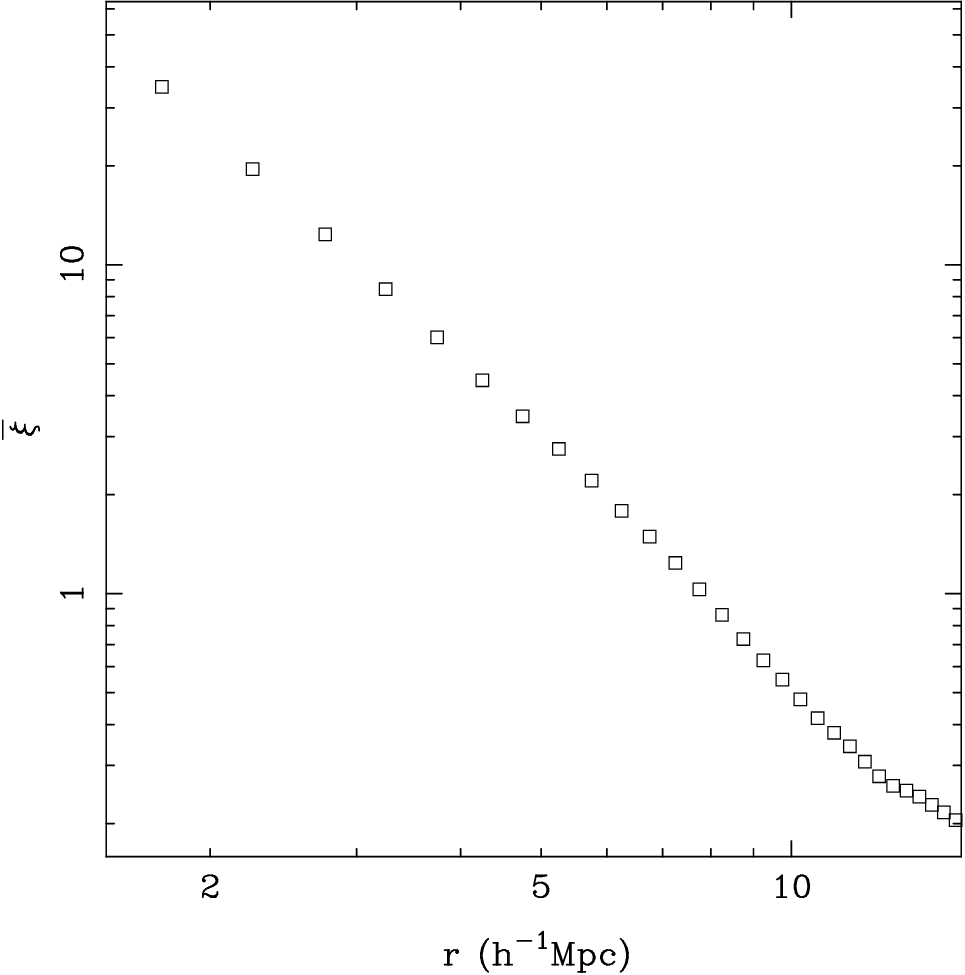}
\caption{This figure shows the averaged two point correlation function
  for halos of mass $M_{halo} \ge 7\times 10^{11}M_\odot$ at $z=3$.
  The x-axis shows the comoving scale in units of h$^{-1}$Mpc.  The
  amplitude of the halo correlation function is comparable to the
  galaxy correlation function at $z=0$, i.e. it is much higher than
  the amplitude of mass correlation function at $z=3$.}
\end{figure}

The large amplitude of correlation function is not completely
surprising.  Analytical models for evolution of bias for halos also
predict that bias increases with increasing redshift (For example, see
Matarrese et al. 1997).  These predictions suggest that the amplitude
of halo correlation function decreases very slowly with increasing
redshift.  However, most of these models do not predict the halo
correlation function at $z=3$ to be as strong as the present day
galaxy correlation function \cite{evolclus}. 

\section{Synthetic Observations}

In this section, we will simulate observations using N-Body
simulations.  We will restrict our study to the CDM class of models.
In particular, we will study clustering
in the standard CDM model ($\Gamma=0.5$) and a variant of SCDM that
reproduces the observed correlation function out to large scales
($\Gamma=0.3$).  These models are normalised so that
$\sigma_8=\sigma(8 h^{-1}Mpc, z=0)=0.6$.  To describe differences
induced by a higher normalisation we will use the same models with
$\sigma_8=1$.   The background cosmology is assumed to be $\Omega=1$,
$H_0=50$km Mpc$^{-1}$ s$^{-1}$.  We will specify all scales in the
comoving coordinates.

The observations indicate a very large scale for the concentration of
galaxies.  In the Einstein deSitter model $\Delta z=0.04$ translates
to $15h^{-1}$Mpc.  The observed angular extent of the structure in
question is greater than or equal to $8' \times 11'$. (One arc minute
corresponds to $0.87h^{-1}$Mpc at $z=3$.)  To study clustering on such
large scales we need to simulate a very large volume and we chose to
work with a simulation box of size $166h^{-1}$Mpc.  All simulations
were done using $128^3$ particles so that mass of each N-Body particle
was approximately $10^{12} M_\odot$.  All N-Body simulations used here
were done using a Particle-Mesh code.  

To simulate ``observations'' of high redshift galaxies we use the
following method.
\begin{itemize}
\item As galaxies form in regions with high density, we begin by
  isolating such regions in the simulation volume. 
\item We project the particles in the selected regions in redshift
  space by assuming that one of the axes is aligned along the line of
  sight.  Velocity along the line of sight and the Hubble redshift
  combine to give the total effective redshift of each particle. 
  \begin{equation} 
    1+z_{tot}=(1+z_{hub}) \; (1+z_{pec}) 
  \end{equation}
  Here $z_{pec} = v_p/c$ with $c$ the speed of light and $v_p$ the
  component of the peculiar velocity of the particle along the line of
  sight.   
\item We then view the selected particles around a large number of
  lines of sight.  For each line of sight all particles within a
  square region of $10' \times 10'$ are included in the field of view.
  The redshift distribution of these particles is then analysed for a
  large number of fields of view. 
\item We smooth the distribution in redshift by using a top hat window
  of width $\Delta z =0.04$.  We search for peaks in the smoothed
  distribution.  As the correspondence between particles and galaxies
  is not very clear, we will measure peaks from the average amplitude
  in units of the standard deviation $\sigma$. 
\end{itemize}

We are interested in clustering at large scales and therefore we have
to use a large simulation volume.  This implies that the mass of
individual particles in the simulation is considerably larger than the
mass associated with the galaxies we are studying.  This clearly makes
it difficult to identify galaxies directly and we can only try to
isolate regions where such galaxies could have formed.  Therefore, we
work with three density thresholds: (1) $\bar\varrho_{c}=0$, i.e. no
cutoff; (2) $\bar\varrho_{c}=2 \varrho_b$, i.e. we will select regions
with $\delta \geq 1$.  Here $\varrho_b$ is the background density and
$\delta$ is the density contrast. And, (3) $\bar\varrho_{c}=4
\varrho_b$, or $\delta \geq 3$.  As we are using very low thresholds,
the frequency of occurrence of large concentrations of galaxies
computed here is essentially a lower limit.

Our aim here is to search for clustering at large scales in redshift
and the width of the structure we hope to reproduce is $\Delta z =0.04$.
However, it is useful to start with smaller bins and average over the
neighbouring bins with a top hat filter.  This ensures that all high
peaks are picked out.  

\begin{figure}
\epsfxsize=3.3truein\epsfbox[46 26 502 494]{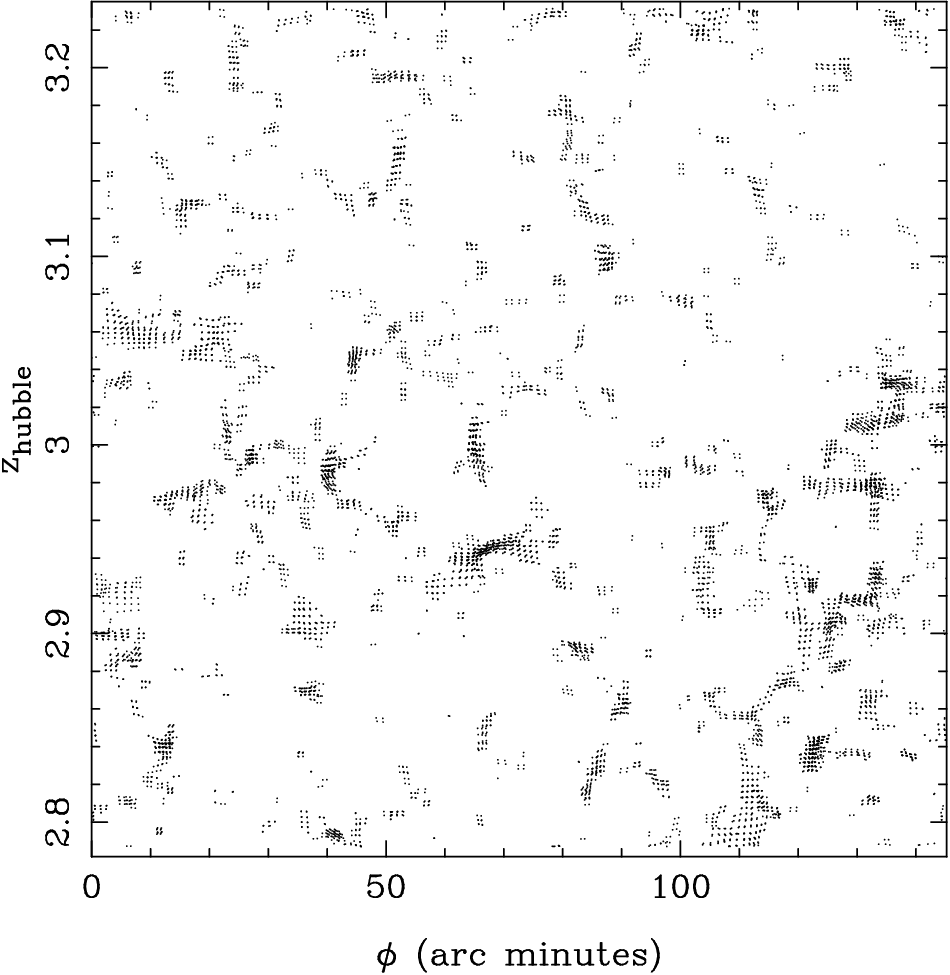}
\epsfxsize=3.3truein\epsfbox[46 26 502 520]{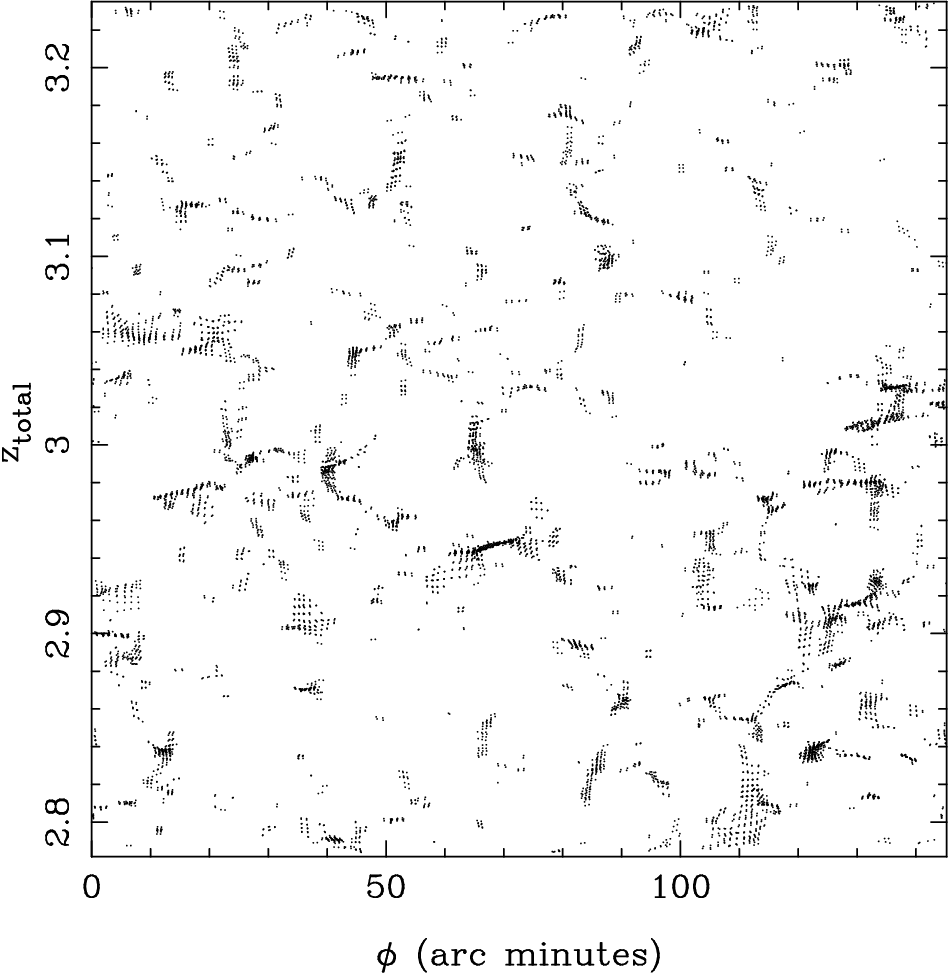}
\caption{This figure shows the dense structures selected using a
  threshold $\bar\varrho_c=4\varrho_b$ in real as well as redshift
  space. The model used in these simulations is CDM with $\Gamma=0.5$,
  normalised to $\sigma_8 = 0.6$.  The figures show projected
  distribution of a slice of thickness $10'$.  The top panel shows the
  particles in real space and for comparison with the second panel and
  observations, we have used the cosmological redshift and angle
  $\phi$ as coordinates.  The lower panel shows the same set of
  particles in redshift space.  Here $1+z_{tot} =
  (1+z_{hub})(1+z_{pec})$, where $z_{pec}=v_p/c$ with $c$ as the speed
  of light and $v_p$ the component of peculiar velocity along the line
  of sight.  It is obvious from this figure that redshift space
  distortions operate on much smaller scales than $\Delta z =0.04$.
  It is also clear that there are clearly some structures with an
  angular extent greater than $10'$, size of the observed
  concentration.}  
\end{figure}

We group the particles in each field of view in bins of
$500$km s$^{-1}$.  We then combined six bins around each of the
smaller bins to get a distribution with $\Delta z = 0.04$ sampled at
every $500$km s$^{-1}$.  We scan this distribution for locating
maxima and in order to avoid over-counting high peaks, we ensure
that a given bin has the largest number of particles as compared to
bins within $\Delta z = 0.02$ on either side.  The amplitude of peaks
is measured from the average in units of standard deviation -- the
average and standard deviation are obtained from a large number of
``fields of view'' through the given simulation box.

\section{Results}

We have plotted a slice from an N-Body simulation in
fig.2.  This figure shows the projection of a slice from a CDM
simulation with $\Gamma=0.5$ and $\sigma_8=0.6$.  The particles shown
here are located in regions with $\delta \geq 3$.  The projection is
shown in both real and redshift space.  For easy comparison with the
relevant observations, we have used angular size and redshift as the
coordinates.  The thickness of this slice is $10'$.  The top panel
shows the distribution of particles in real space and the lower panel
shows the same set of particles in redshift space.  It is clear that
redshift space distortion tends to squeeze many structures into a thin
sheet perpendicular to the line of sight. However, the scale where
this effect is important is much smaller than $\Delta z = 0.04$.
The sizes of largest structures -- angular sizes as well as
the extent in redshift -- and the effects of redshift space distortion
can be seen clearly.  Redshift space distortions play a very important
role at small scales (McGill 1990a, 1990b) but for the scales that we
are interested in, i.e. $\Delta z =0.04$, these effects are smoothed
out.  At these scales, the frequency with which such structures appear
is identical in the real and redshift space.

Fig.3 shows the redshift distribution in two fields of view obtained
using the method outlined in \S{3}.  These particular examples were
chosen to exhibit the variety of distributions and peaks seen in such
systems.  In one case we see a semi-periodic set of peaks whereas in
the other case we see only one isolated peak. 

Table 1 lists the frequency of occurrence of $N\sigma$ peaks above
average in the redshift space distribution.  Here we have extrapolated
from the range of redshift covered by the simulation to that used in
Steidel et al. (1997) to facilitate comparison.  We can summarise the
conclusions as follows:
\begin{itemize}
\item Models with same normalisation predict similar frequency of
  occurrence for peaks with $N\sigma$ more particles than average.
  This is to be expected as both models have similar power at the
  relevant scale.
\item Models with higher normalisation predict a higher frequency of
  occurrence for the unbiased case.
\item Models with higher normalisation predict a lower frequency of
  occurrence for $5\sigma$ peaks above average for the high density
  threshold.  This can be understood if the statistical bias depends
  only on $\nu$ defined in the introduction.  This is small for models
  with the higher normalisation, and hence we do not see the effect
  of strong clustering of rare peaks. 
\end{itemize}

\begin{figure}
\epsfxsize=3.3truein\epsfbox[44 31 533 530]{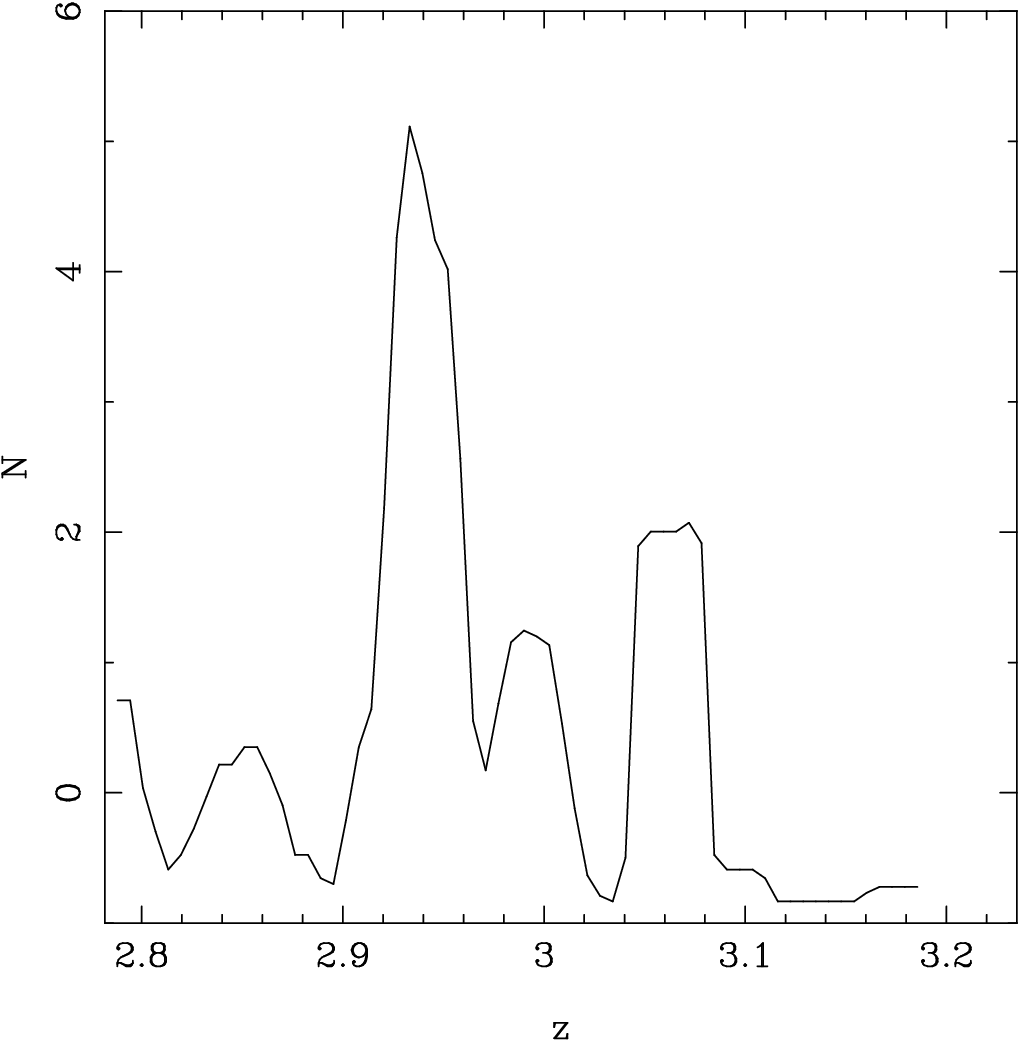}
\epsfxsize=3.3truein\epsfbox[44 31 533 560]{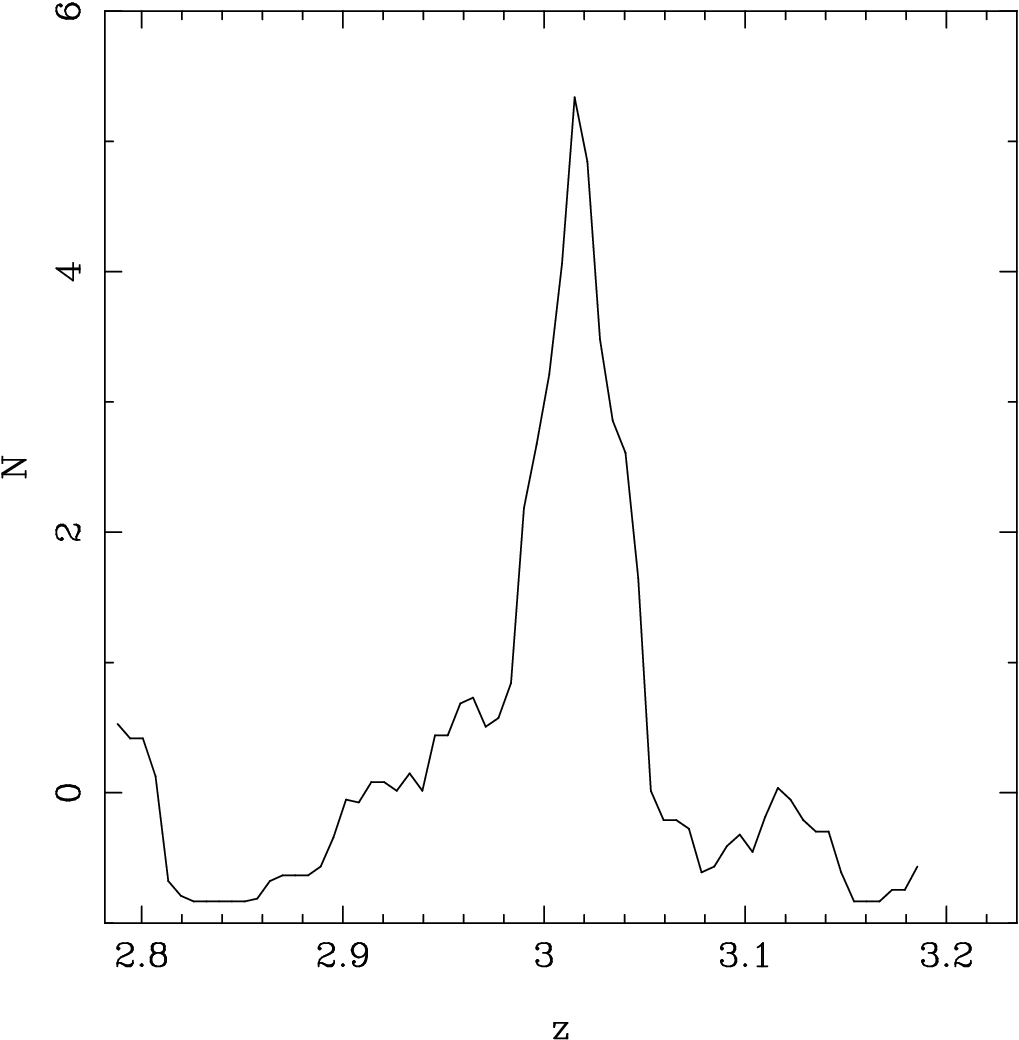}
\caption{Distribution of objects in redshift space.  These panels
  show the redshift distribution in two $10' \times 10'$ fields of
  view through a simulation volume.  These are from the simulation 
  used in fig.2.  These frames show a few high peaks in the
  distribution.  The y axis is the height of a peak measured from the
  average in units of standard deviation $\sigma$.  The curve shows an
  average over $\Delta z=0.04$ sampled at intervals of $\Delta
  z=0.0067$.}
\end{figure}

\begin{table}
\begin{tabular}{llllll}
$\Gamma$ \ \ \ \ & $\sigma_8$ \ \ \ \ & $\delta_c$ \ \ \ \ & $N=3$ \ \
\ \ & $N=4$ \ \ \ \ & $N=5$ \ \ \ \ \\
0.5 & 0.6 & $-$ & 0.45 & 0.07 & 0.01 \\
0.5 & 0.6 & $1$ & 0.64 & 0.11 & 0.01 \\
0.5 & 0.6 & $3$ & 1.01 & 0.45 & 0.18 \\
0.3 & 0.6 & $-$ & 0.46 & 0.07 & 0.01 \\
0.3 & 0.6 & $1$ & 0.77 & 0.19 & 0.04 \\
0.3 & 0.6 & $3$ & 0.87 & 0.51 & 0.24 \\
0.5 & 1.0 & $-$ & 0.71 & 0.16 & 0.03 \\
0.5 & 1.0 & $1$ & 0.72 & 0.18 & 0.04 \\
0.5 & 1.0 & $3$ & 0.99 & 0.34 & 0.12 \\
0.3 & 1.0 & $-$ & 0.73 & 0.16 & 0.03 \\
0.3 & 1.0 & $1$ & 0.73 & 0.17 & 0.04 \\
0.3 & 1.0 & $3$ & 1.00 & 0.34 & 0.12 \\
\end{tabular}
\caption{This table lists the frequency of $N\sigma$ peaks above
  average in the models of structure formation studied here.  The
 first two columns refer to the parameters of the model, the shape
 parameter $\Gamma$ and normalisation $\sigma_8$.  The third column 
 lists the threshold in density contrast used to select particles.
 The last three columns list the frequency for finding $3$, $4$ and
 $5$ $\sigma$ peaks above average in fields of view of $10' \times
 10'$ and redshift range $2 \le z \le 3.5$.}
\end{table}

\section{Discussion}

We have shown that the observations of strong clustering of galaxies
at high redshifts are not in conflict with the popular models of
structure formation.  We have demonstrated this in two different
ways: first by showing that the bias can indeed be very large at early
times.  We have also generated synthetic observations in a few models
and have shown that the frequency of occurrence of large concentrations
of galaxies is compatible with observations. 

Fig.2 suggests that still larger structures maybe found in future
searches.  The distribution of sizes of largest structures in surveys
at high redshift -- both angular size and the extent in redshift --
may be used to discriminate between different models.

Strong clustering of galaxies at high redshifts implies that the
galaxy clustering evolves in a very different manner as compared to
clustering in the underlying mass distribution.  A detailed study of
the evolution of halo clustering and its implications for galaxy
clustering has been presented elsewhere \cite{evolclus}.

\section*{ACKNOWLEDGEMENT}

I thank Max Pettini for drawing my attention to this observation.  I
acknowledge the support of PPARC fellowship at the Institute of
Astronomy.

\label{lastpage}


\begin{thebibliography}{}

\bibitem[\protect\citename{Bagla} 1997]{evolclus} Bagla J.S. 1997,
  astro-ph/9711081, Submitted to MNRAS

\bibitem[\protect\citename{Bardeen et. al.} 1986]{bbks} Bardeen J.M.,
Bond J.R., Kaiser N. and Szalay A.S. 1986, ApJ 304, 15 

\bibitem[\protect\citename{Efstathiou, Bond and White} 1992]{ebw}
Efstathiou G., Bond J.R. and White S.D.M. 1992, MNRAS 258, 1p

\bibitem[\protect\citename{Gunn and Gott} 1972]{gunngott} Gunn
J.E. and Gott J.R. 1972, ApJ 176, 1 

\bibitem[\protect\citename{Matarrese et al.} 1997]{manybias} Matarrese
  S., Coles P., Lucchin F. and Moscardini L. 1997, MNRAS 286, 115

\bibitem[\protect\citename{McGill} 1990a]{red1} McGill C. 1990a, MNRAS
  242, 428

\bibitem[\protect\citename{McGill} 1990b]{red2} McGill C. 1990b, MNRAS
  242, 544

\bibitem[\protect\citename{Press and Schechter} 1975]{ps75} Press
  W.H. and Schechter P. 1975, ApJ 187, 452 

\bibitem[\protect\citename{Steidel et al} 1997]{wallobs} Steidel C.C.,
Adelberger K.L., Dickinson M., Giavalisco M., Pettini M. and Kellogg
M. 1997, Submitted to ApJ.

\end{thebibliography}
\end{document}